# Can a Wi-Fi WLAN Support a First Person Shooter?

Jose Saldana, Juan Luis de la Cruz, Luis Sequeira, Julián Fernández-Navajas, José Ruiz-Mas
CeNITEQ Group, Aragon Institute of Engineering Research (I3A)
EINA, University of Zaragoza
Zaragoza, Spain
{jsaldana, jlcruz, sequeira, navajas, jruiz}@unizar.es

*Abstract*—In corporate and commercial environments, the deployment of a set of coordinated Wi-Fi APs is becoming a common solution to provide Internet coverage to moving users. In these scenarios, real-time services as online games can also be present. This paper presents a set of experiments developed in a test scenario where an end device moves between different APs while generating game traffic. A WLAN solution based on virtual APs is used, in order to make the handoffs transparent for Layer 3. The results show that it is possible to maintain an acceptable level of subjective quality during the handoff. At the same time, it is set clear that the fact of having a gamer in an AP could be taken into account by radio resource management algorithms, in order to provide a better quality.

## I. Introduction

Online games with tight real-time constraints are becoming ubiquitous: they are no longer exclusive of high-end PCs, but many of them have been ported to tablets or even smartphones. Whereas the user mobility with a laptop can be considered as *nomadic* (i.e. a user may move, but he/she will stay for a long time in the same place), smartphone and tablet users may walk while using real-time services.

The Wi-5 *(What to do With the Wi-Fi Wild West)* Project is exploring a set of functionalities to be included in a pool of coordinated smart Wi-Fi APs. One of the aims of the Project is to make it possible for these APs to support real-time games with quality. This includes resource management algorithms that take into account the nature of each flow, and its coexistence with other services. In addition, seamless handovers between APs are not only required for supporting user mobility, but also for assigning a static player to another AP, if it is required for the optimization of radio resources.

Once a testbed including the basic functionalities has been built, this paper presents some tests ran with the aim of answering these two research questions:

- Can this Wi-Fi WLAN support seamless handovers between different APs? In this case, "seamless" means that the player of a First Person Shooter, usually considered the game genre with the tightest real-time constraints, must experience a good quality.

- Should the fact of having a player in an AP be considered as an input for the resource management algorithm?

Answering these two questions will help us define the use cases, the requirements, and the system architecture for a mature Wi-Fi solution. The remainder of the paper is as follows: The architecture is summarized in Section II; Section III presents the test setup and the results, and the paper ends with the Conclusions and Future Work section.

## II. Architecture of the Wi-Fi WLAN

The architecture being developed by the Wi-5 project not only includes Wi-Fi APs, but also considers the possibility of coordination with 3G/4G mobile networks for offloading purposes. However, for the sake of simplicity, in this paper we will only consider a scenario including a number of Wi-Fi APs controlled by a single entity (as it may happen in a mall, an airport or a business center).

As shown in Fig. 1, a number of APs are connected to a central controller using a wired network as a backhaul. We have selected Odin [1], and open-source solution able to provide orchestration in WLAN scenarios. The main reason for selecting Odin is its suitability for supporting load balancing, seamless handover and other interesting features. In regular Wi-Fi, the handoff is usually triggered by the user device, so it is not controlled by the network, and it may require 1 or 2 seconds [2]. In contrast, the selected solution allows the network to control the mobility and to select the best moment for the handoff. In the remainder of this section we will briefly summarize the main characteristics of Odin.

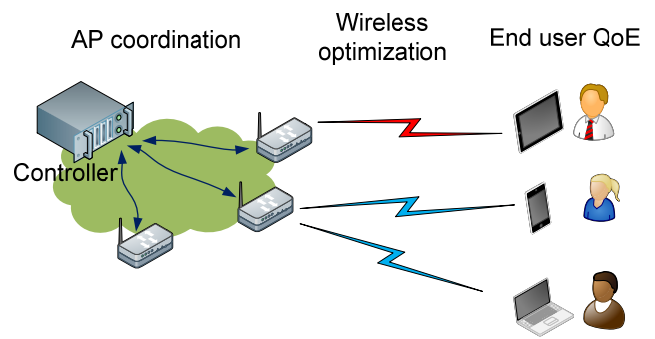

Fig. 1. Basic scheme of the smart APs architecture

This work has been partially financed by the EU H2020 Wi-5 project (Grant Agreement no: 644262), and European Social Fund in collaboration with the Government of Aragon.

The system is based on commodity OpenWRT APs with a single radio. Each AP runs Click Modular Router [3], which enables the possibility of directly managing the traffic. In addition, Open vSwitch[1] is installed, thus making the internal switch of each AP behave as an Openflow switch. The controller runs Floodlight Openflow Controller[2] in order to manage all the switches of the APs. The resource management algorithms are added as applications on top of the controller.

Odin uses the LVAP (Light Virtual Access Point) abstraction [1], which is created by the controller for each user terminal (client), making it see the whole WLAN as a single AP. For that aim, the AP, instead of sending the frames with its own MAC address, uses a MAC specifically created for the client. If the client moves to another AP, the controller moves the LVAP accordingly, so the user terminal does not notice the AP change. This is illustrated in Fig. 2: first, a "Subscription" is added to each AP (1). When a client moves (2), the destination AP detects a radio signal power increase, and publishes a message (3). The controller may decide to handoff the client, and the LVAP accordingly (4).

All in all, the use of LVAPs makes it possible to hide to the terminal the switch between APs, thus avoiding the delay produced by re-association. In fact, the handover is totally transparent for Layer 3.

## III. TESTS AND RESULTS

Once we have summarized the Odin solution, in this section we present some tests we have run in order to answer the research questions presented in the Introduction. The objectives of Wi-5 include the support of real-time services with good quality, and First Person Shooters represent a good example of a service with tight real-time requirements [4].

### A. Description of the Testbed

We have implemented Odin in a controlled laboratory environment, using the publicly available implementation[3]. These are the characteristics of the machines:

- Two TP-Link1043NDv2 APs are used, both configured in channel 6 (2.4 GHz band).
- The controller is a commodity PC running Linux Debian 6.
- The game traffic is generated by another Linux machine, using D-ITG traffic generator [5], which includes an option for generating *Quake3* (a popular First Person Shooter) traffic.

### B. Results

In the first test, a single machine is connected to the Odin Wi-Fi SSID, and it generates a *Quake3* traffic flow. No other machines are connected to that SSID. The mobility application has been configured to handoff the client every 3 seconds from one AP to another. Fig. 3 presents the delay of each packet during 24 seconds (8 handoffs). It can be observed that the delay does not significantly increase after the handoff, and it is usually below 15 ms. The jitter (delay standard deviation) is 5.5 ms. The packet loss rate is 3.25%.

A subjective quality estimator [6], based on delay and jitter, has been used to obtain the results shown in Table I. Different values of Round Trip Time (typical intra and inter-region values) have been added, in order to consider the time required to reach the game server, taking into account that in our setup all the machines are in the same LAN. The value of MOS (Mean Opinion Score) ranges from 1 *(bad quality)* to 5 *(excellent)*. If the MOS scale of VoIP is used, it is considered acceptable above 3.5. However, some works consider that a value of 3 can be good [7], but gamers will exchange to another server if MOS is roughly 2. In our case, MOS values above 3 can be obtained.

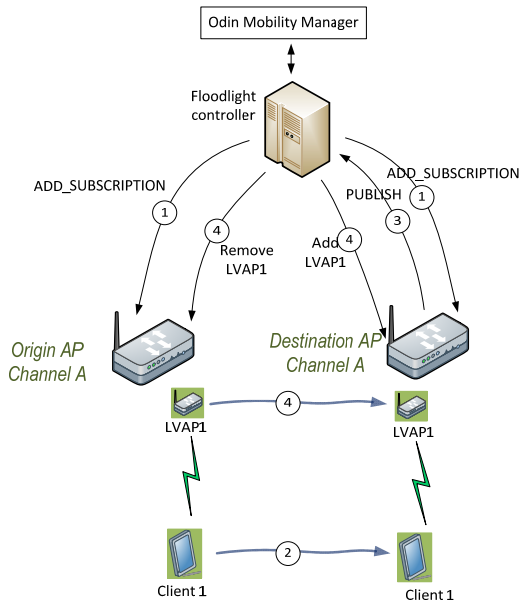

Fig. 2. Scheme of a handover with Odin

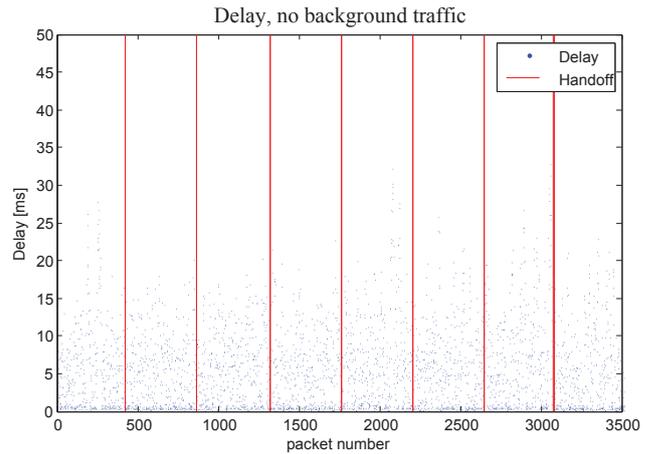

Fig. 3. Delay *Quake 3* packets when no background traffic is present

---

[1] Open v Switch, http://openvswitch.org/

[2] Floodlight SDN Controller: http://www.projectfloodlight.org/floodlight/

[3] https://github.com/lalithsuresh/odin

TABLE I.  SUBJECTIVE QUALITY ESTIMATION DEPENDING ON NETWORK DELAY

| Scenario | G-Model Subjective Quality Estimator[a] | | |
|---|---|---|---|
| | *Delay (Round Trip Time)* | *Jitter* | *MOS* |
| LAN | 5 ms | 5.5 ms | 3.73 |
| Intra-region | 20 ms | 5.5 ms | 3.58 |
| Inter-region | 80 ms | 5.5 ms | 3.04 |

[a.] Please note that this model was developed for *Quake 4*, but we are using it for *Quake 3*, so the results are only estimative

The second test (Fig. 4) is similar, but another machine is downloading a big file (a Debian ISO image) from the Internet using the Odin SSID. It can be observed that the delay increases significantly (following a pattern similar to the sawtooth TCP rate increase of the FTP download), making the game unplayable. In addition, many packets are lost. There are some moments where a correlation between handoffs and delay is observed (see e.g. a delay increase after the handoff corresponding to packet number 1740), but the delay increase caused by the handoff is not significant. Long bursts of lost packets appear.

We will now try to answer the two research questions raised in the Introduction. First of all, we can conclude that the solution based on LVAPs [1] is able to support seamless handovers of an online game traffic. It has been shown that the delay added to the packets after a handover is not significant.

Regarding the second question, the results set clear that the coexistence of an online game and an FTP download produces a significant delay increase on the real-time application. Therefore, if the resource management algorithms were aware of the presence of a real-time flow in an AP, they would be able to distribute the terminals in a way that avoids this coexistence. This result encourages us to include monitoring tools able to detect real-time flows (e.g. [8]) in the architecture, and to consider this factor as an important input for the resource management algorithms.

## IV. CONCLUSIONS AND FUTURE WORK

This paper has presented a set of experiments ran in a scenario where an end device moves between different APs while generating game traffic. A WLAN solution based on virtual APs has been used in order to make the handoffs transparent for Layer 3. The results show that it is possible to maintain an acceptable subjective quality level during the handoff. At the same time, it is set clear that the fact of having a gamer in the AP could be taken into account by the radio resource management algorithms, in order to provide a better quality: the results have shown that the coexistence of online games' traffic and FTP in the same AP may harm the QoE of the player, so perhaps these flows should be attended by different APs.

As future work, our plan is to integrate the seamless handover with the channel switch. For that aim, the channel switch and the handover have to be synchronized. This would allow the use of different channels in the WLAN, thus reducing the interference level.

Another objective is to tune the handover procedure in order to adapt it to the services currently running in the client. For example, the current implementation first adds the LVAP to the destination AP, and then removes it from the origin, so this may produce some duplicated packets. Depending on the service, it may be preferable to do the opposite (removing the LVAP first), taking into account that some services have a good tolerance to packet loss. For example, in [6] it was shown that players of *Quake 4* did not notice packet loss up to 35%.

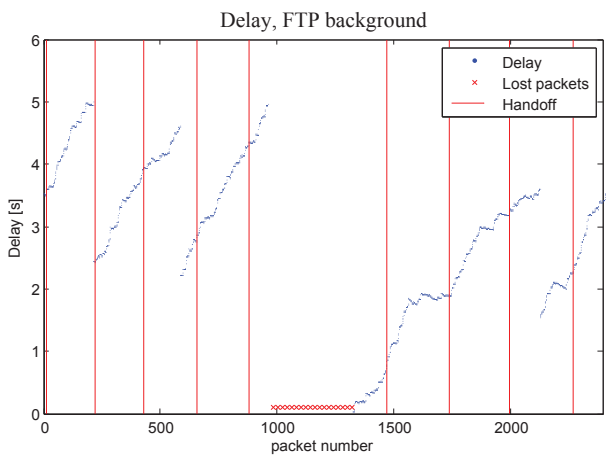

Fig. 4. Delay *Quake 3* packets when coexisting with a FTP background flow